\documentclass{ifacconf}

\makeatletter
\let\old@ssect\@ssect 
\makeatother

\usepackage{graphicx}    
\usepackage{color,array}
\usepackage{xcolor}
\usepackage[colorlinks,urlcolor=blue,linkcolor=blue,citecolor=blue]{hyperref}

\usepackage{amssymb}
\usepackage{amsmath}
\usepackage{amsfonts}
\usepackage{cite}
\usepackage{siunitx}
\usepackage{bm}
\usepackage{algorithm}
\usepackage{algpseudocode}
\usepackage{longtable,tabularx}
\usepackage{multirow}
\usepackage{subfigure}
\usepackage{placeins}
\usepackage{float}
\usepackage{nomencl}
\usepackage{booktabs}
\usepackage{natbib} 

\makeatletter
\def\@ssect#1#2#3#4#5#6{%
  \NR@gettitle{#6}
  \old@ssect{#1}{#2}{#3}{#4}{#5}{#6}
}
\makeatother

\newcommand{\YL}[1]{{ #1}}
\newcommand{\YLTR}[1]{{ #1}}
\newcommand{\YLMS}[1]{{ #1}}

\def\be{\begin{equation}}
\def\ee{\end{equation}}

\begin{document}
\begin{frontmatter}

\title{Physics-Guided State-Space Model Augmentation Using Weighted Regularized Neural Networks	\thanksref{footnoteinfo}} 

\thanks[footnoteinfo]{This work is funded by the European Union (ERC, COMPLETE, 101075836). Views and opinions expressed are however those of the author(s) only and do not necessarily reflect those of the European Union or the European Research Council Executive Agency. Neither the European Union nor the granting authority can be held responsible for them.}

\author[First]{Yuhan Liu} 
\author[First,Second]{Roland Tóth} 
\author[First]{Maarten Schoukens}

\address[First]{Control Systems Group, Eindhoven University of Technology,
Eindhoven, the Netherlands}
\address[Second]{Systems and Control Laboratory, Institute for Computer Science
and Control, Budapest, Hungary}

\begin{abstract} Physics-guided neural networks (PGNN) is an effective tool that combines the benefits of data-driven modeling with the interpretability and generalization of underlying physical information. However, for a classical PGNN, the penalization of the physics-guided part is at the output level, which leads to a conservative result as systems with highly similar state-transition functions, i.e. only slight differences in parameters, can have significantly different time-series outputs. Furthermore, the classical PGNN cost function regularizes the model estimate over the entire state space with a constant trade-off hyperparameter. In this paper, we introduce a novel model augmentation strategy for nonlinear state-space model identification based on PGNN, using a weighted function regularization (W-PGNN). The proposed approach can efficiently augment the prior physics-based state-space models based on measurement data. A new weighted regularization term is added to the cost function to penalize the difference between the state and output function of the baseline physics-based and final identified model. This ensures the estimated model follows the baseline physics model functions in regions where the data has low information content, while placing greater trust in the data when a high informativity is present. The effectiveness of the proposed strategy over the current PGNN method is demonstrated on a benchmark example.


\end{abstract}


\end{frontmatter}

\section{Introduction}
\label{sec:1}
Model-based design plays a crucial role in achieving satisfactory performance for complex dynamic systems by providing an interpretable framework that facilitates a deep understanding of system behaviors,  including nonlinearities such as damping and friction.
However, the accurate and comprehensive system dynamics that can be modeled by first principle laws are often costly to obtain. 

Nonlinear system identification \citep{schoukens2019nonlinear} is a well-established topic and can be characterized by a wide range of model classes such as state-space models \citep{schon2011system}, block-oriented models \citep{schoukens2017identification}, NARMAX \citep{billings2013nonlinear}, etc. 
Wherein, extensive research \citep{verdult2002non,paduart2010identification,schon2018probabilistic} on identification with \emph{nonlinear state-space} (NLSS) models has shown its flexibility for handling multi-variable systems with potentially fewer parameters. \YLTR{Estimation of state-space models is advantageous for the subsequent control design, given the dependency of many nonlinear control methods on such representation of the system behavior}.

\YLTR{\emph{Artificial neural networks} (ANNs) have long been a focus of interest in the field of nonlinear system identification because of their high expressiveness, flexibility, and capability of approximating functions with arbitrary accuracy \citep{scarselli1998universal}. 
In \cite{suykens1995nonlinear}, recurrent neural networks have already been employed to represent a nonlinear state-space model. This structure is referred to as \emph{state-space neural network} (SS-NN) and has been further discussed in
\citep{amoura2011state,forgione2020model,schoukens2021improved,beintema2023deep}. 
Recently, \cite{beintema2023deep} have introduced a computationally efficient nonlinear system state-space identification algorithm based on a subspace-encoder network (SUBNET).} Nonetheless, \YLTR{ANNs are typically black-box models} that lack physical interpretation, and exhibit poor generalization capabilities outside the training dataset, especially when the training data is limited.
\YLTR{Hence, even though the ANNs may exhibit improved accuracy compared with first-principle modeling, deploying such models in practice or the controllers that are designed for them is simply dangerous.}

To address this issue, \emph{physics-guided neural network} (PGNN) \citep{karpatne2017physics} has been introduced, also within the field of systems and control \citep{bolderman2024physics}, that \YLTR{ensures the interpretability and generalization capabilities of the estimated models.}
Compared with the ANNs, a physics-based cost function is incorporated into the optimization objective of PGNN, ensuring that the learned model not only achieves high accuracy on the training dataset, but also shows consistency with known physics laws on the unseen region without the need for large amounts of ground truth data. 

However, there are some open technical issues with using  PGNNs in nonlinear state space model identification. \YLMS{First, the classical PGNN does not perform the model augmentation, i.e., the prior model is only used to compute the physics-based regularization term in the cost function.}
Second, the classical PGNN penalizes the difference between the physics model and the identified model at the output level, which can lead to conservative estimation results. \YLTR{This is because systems with highly similar state-transition functions can have significantly different time-series outputs.} Furthermore, the physics-based term of the classical PGNN regularizes the model difference over the entire state space, which makes this approach lose some flexibility, especially when the assumed prior model is \YLTR{inaccurate}.


Motivated by these facts, this paper proposes an innovative PGNN-based state-space modeling strategy for nonlinear system identification, namely, W-PGNN,  to efficiently complete prior physics-based state-space models with a weighted regularized SS-NN. 
The main contributions are as follows:

 1) \YLMS{A new weighted-regularization cost function is designed to penalize the difference between both the state and output functions of the baseline physics-based and final identified models in regions where measured data provides little information.}

2) \YLTR{Compared to the classical PGNNs, the proposed identification approach makes more extensive use of the pre-existing approximate model. The learned dynamics are capable of adhering to the data in regions with high information content, and preserving the behavior of the baseline physics model outside this region.}
This significantly enhances the flexibility of the SS-NN model.

The remainder of this paper is organized as follows. Section \ref{sec:2} introduces the nonlinear model class and the identification method with a state-space neural network. The classical PGNN method is discussed in Section \ref{sec:3}. The proposed W-PGNN method is detailed in Section \ref{sec:4}.
Numerical simulation results are provided in Section \ref{sec:5}, followed by the conclusions in Section \ref{sec:6}.

\emph{Notation}: $\mathbb{R}$ and $\mathbb{Z}$ denote the sets of real numbers and integers, respectively. The 2-norm of a vector or a matrix is denoted as $\|\cdot\|$.
$\mathrm{vec}({x}_1,...,{x}_n)=[{x}_1^{\top} \ \cdots \ {x}_n^{\top}]^{\top}$ denotes the column-wise composition of vectors.
$\mathcal{N}(0,1)$ is the standard normal distribution, while
$\mathcal{U}(a,b)$ represents a uniform distribution with a support from $a$ to $b$.

\section{Problem Statement}
\vspace{-1mm}
\label{sec:2}
\subsection{Nonlinear Model Class}
\vspace{-1mm}
Consider the following discrete-time state-space model as the data-generating system: \vspace{-0.1mm}
\be
\label{1}
\begin{aligned}
x(k+1) & =f(x(k), u(k)), \\
y_0(k) & =g(x(k), u(k)),
\end{aligned}
\ee
where $u(k)\in\mathbb{R}^{n_\mathrm{u}}$ denotes the input, $x(k)\in\mathbb{R}^{n_\mathrm{x}}$ is the state, $y(k)\in\mathbb{R}^{n_\mathrm{y}}$ is the output, and $k\in\mathbb{Z}$ represents the discrete time. Additionally, ${f}:\mathbb{R}^{n_\mathrm{x}\times n_\mathrm{u} } \to \mathbb{R}^{n_\mathrm{x}}$ and ${g}:\mathbb{R}^{n_\mathrm{x}\times n_\mathrm{u} } \to \mathbb{R}^{n_\mathrm{y}}$ are bounded deterministic vector functions. The training dataset $\mathcal{D}= { \{({y}(k),{u}(k)) \}_{k=1}^N}$ contains $N$ noisy outputs $y(k)=y_0(k)+v(k)$, collected from an experiment on \eqref{1}, where the noise $v(k)$ is assumed to be a zero-mean random signal with finite variance independent from the input $u(k)$.

Assume that we only have access to an a priori known state-space model:
\be
\label{2a}
\begin{aligned}
\tilde{x}(k+1) & =\tilde{f}\left(\tilde{x}(k), u(k)\right), \\
\tilde{y}_0(k) & =\tilde{g}\left(\tilde{x}(k), u(k)\right),
\end{aligned}
\ee
with the state and output $\tilde{x}(k)\in\mathbb{R}^{n_\mathrm{x}}$ and $\tilde{y}(k)\in\mathbb{R}^{n_\mathrm{y}}$ that has the same model order as \eqref{1}. Note that the functions $\tilde{f}(\cdot)$ and $\tilde{g}(\cdot)$ constitute the physically well-interpretable and a priori know dynamics of the system \eqref{1}, i.e., the nominal model.
However,  the prior model \eqref{2a} does not accurately capture the true dynamics \eqref{1}. For instance, there may exist local nonlinearities in certain regions, which are not able to be obtained by a rough identification or modeling based on first principles.
Hence, it is essential to augment this a priori known model using newly measured data through nonlinear system identification.

\subsection{State-Space Neural Network  Identification }

To this end, we consider the following nonlinear discrete-time state-space model of \eqref{1}, which has the following structure:
\be
\label{2b}
\begin{aligned}
\hat{x}(k+1) & =\tilde{f}\left(\hat{x}(k), u(k)\right)+f_{\theta}\left(\hat{x}(k), u(k)\right), \\
\hat{y}(k) & =\tilde{g}\left(\hat{x}(k), u(k)\right)+g_{\theta}\left(\hat{x}(k), u(k)\right),
\end{aligned}
\ee
where $f_{\theta}(\cdot)$ and $g_{\theta}(\cdot)$ are the \emph{completion functions} that model the dynamics that
cannot be captured reliably by the idealistic model \eqref{2a},
and are represented by fully connected feedforward neural networks with one hidden layer containing $n_n$ neurons and a linear output layer:
\be
f_{\theta}(x(k),u(k))=W_x \phi\left(\left[W_{f x} W_{f u}\right]\left[\begin{array}{l}
x(k) \\
u(k)
\end{array}\right]+b_f\right)+b_x
\ee
where $\phi(\cdot)\in\mathbb{R}^{n_\mathrm{u}\times 1 }$ denotes the activation function, $W_x$, $W_{fx}$, $W_{fu}$, $b_f$ and $b_x$ represent the weight and bias parameters of the neural network with proper dimensions, respectively. 
A similar representation is used for $g_{\theta}(x(k),u(k))$.

As discussed in \cite{suykens1995nonlinear} and \cite{schoukens2021improved}, the state-space model \eqref{2b} can be written in a specific form of a recurrent neural network, i.e., an SS-NN. The parameters $\theta$ for the SS-NN can be trained by optimizing the data-based cost function over $N$ samples:
\be
\begin{aligned}
\label{cost_ANN}
V_{\mathrm{Data}}(\theta,N) &= \frac{1}{N}\sum_{k=1}^{N}\left\|y(k)-\hat{y}(k |\theta)\right\|^2 \\
\hat{\theta}&=\arg \min _\theta V_{\mathrm{Data}}(\theta,N)
\end{aligned}
\ee
where $\hat{y}(k|\theta)$ is the simulated output of the model \eqref{2b} given the parameter vector $\theta$. More detailed discussions of SS-NN are provided in Section \ref{sec:4}.


From \eqref{cost_ANN}, it is obvious that the ANN simply learns the mapping between system input and output data without considering any prior knowledge about the underlying physics. This makes it difficult for ANN  to have good generalization performance outside of the training region, especially when the dataset is limited. 

\section{Classical PGNN for System Identification}
\label{sec:3}
In this section, we briefly introduce the concept of classical PGNN. Compared with the baseline ANN approach, there is an additional regularization term in the cost function to force the learnt model to follow the prior model even outside the training region.

The classical PGNN is trained by minimizing the following cost function:
\be
\begin{aligned}
\label{4}
V(\theta,N,\bar{N}) &=V_{\mathrm{Data}}(\theta,N) + \gamma V_{\mathrm{Phy}}(\theta,\bar{N})\\
\hat{\theta}&=\arg \min _\theta V(\theta,N,\bar{N})
\end{aligned}
\ee
where $V_{\mathrm{Data}}(\theta,N)$ is given by \eqref{cost_ANN}, and the physics-based penalized term $V_{\mathrm{Phy}}(\theta,\bar{N})$ is given by:
\be
V_{\mathrm{Phy}}(\theta,\bar{N}) =  \frac{1}{\bar{N}}\sum_{k=1}^{\bar{N}}\left\|\tilde{\bar{y}}(k)-\hat{\bar{y}}(k |\theta)\right\|^2
\ee
where $\tilde{\bar{y}}(k)$ and $\hat{\bar{y}}(k|\theta)$ are the output response of the a priori known model \eqref{2a} and the simulated model \eqref{2b} given the regularization input signal $\bar{u}(k)$, respectively,
and $\gamma\in\mathbb{R}_{>0}$ is the constant trade-off hyperparameter \YLTR{that balances between the data-fitting term and the regularization term in the overall cost. }
In this way, the prior model \eqref{2a} is embedded in the trained ANN. \YLMS{Note that the physics-based cost $V_{\mathrm{Phy}}(\theta,\bar{N})$ does not rely on the measurement $y(k)$ from system \eqref{1}. It is evaluated over a separate regularization dataset $\mathcal{D}_{\mathrm{Reg}}=\{\tilde{\bar{y}}(k),\bar{u}(k)\}_{k=1}^{\bar{N}}$ generated by the user using the baseline physics model \eqref{2a}.}

\begin{figure} 
\centering\includegraphics[width= 3.5in]{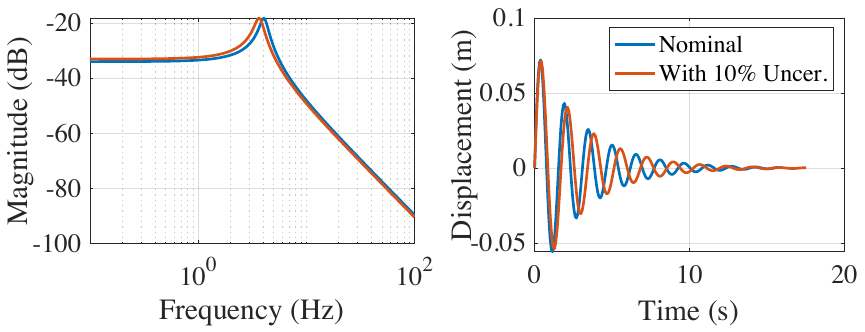}
\vspace{-2mm}
\caption{Frequency and time responses of two mass spring damper systems with only $10\%$ uncertainties in the system parameters. It can be observed that with a slight shift of resonance frequencies, the outputs are significantly different.}
\label{fig:1}
\end{figure}

As can be seen in \eqref{4}, the penalization of the physics-guided part is at the output level. However, this can lead to a conservative result as systems with highly similar state-transition functions, i.e. only slight differences in parameters, can have significantly different time-series outputs. A good example of this is given by two mass spring damper systems with slightly different resonance frequencies, as shown in Fig.~\ref{fig:1}. Furthermore, the classical PGNN cost function regularizes the model estimate for the whole state space, which means that the a priori known model is assumed to hold equally for any unseen region. Though this feature enables the classic PGNN to have better generalization performance than the baseline NN, but in an ideal setting, we would like to trust the information in the data when a high informativity is present, while we would like to follow the prior model in regions where the data provides little information, i.e., to preserve the behavior of the a priori known model. 

\section{Weighted PGNN Method}
\label{sec:4}
\subsection{Weighted function regularization}
\YLMS{Unlike other model augmentation strategies, for instance, \citep{hoekstra2024learning}, our approach aims to regularize state-space neural network estimation using a reference model and penalize the difference between physics and identified model at both the state and output levels. Moreover, the regularization should only be active in the regions where no data is present, i.e. the reference model prescribes the dynamics that the learned model should fall back to outside the training area.}

The proposed approach starts by generating a surrogate input sequence $\bar{u}$ of length $\bar{N}$. It is worth noting that $\bar{u}$ is not applied to the true system during optimization to acquire output measurements, but plays a role in the regularization of the proposed approach. 
\YL{Ideally, $\bar{u}$ should cover the full range of operation of the system.}
Then, the model estimate is evaluated both applying $u$ and $\bar{u}$ on system \eqref{2b}, where the second input sequence results in the state sequence $\hat{\bar{x}}$.

\begin{figure*}[htbp] 
\centering\includegraphics[width= 5.5in]{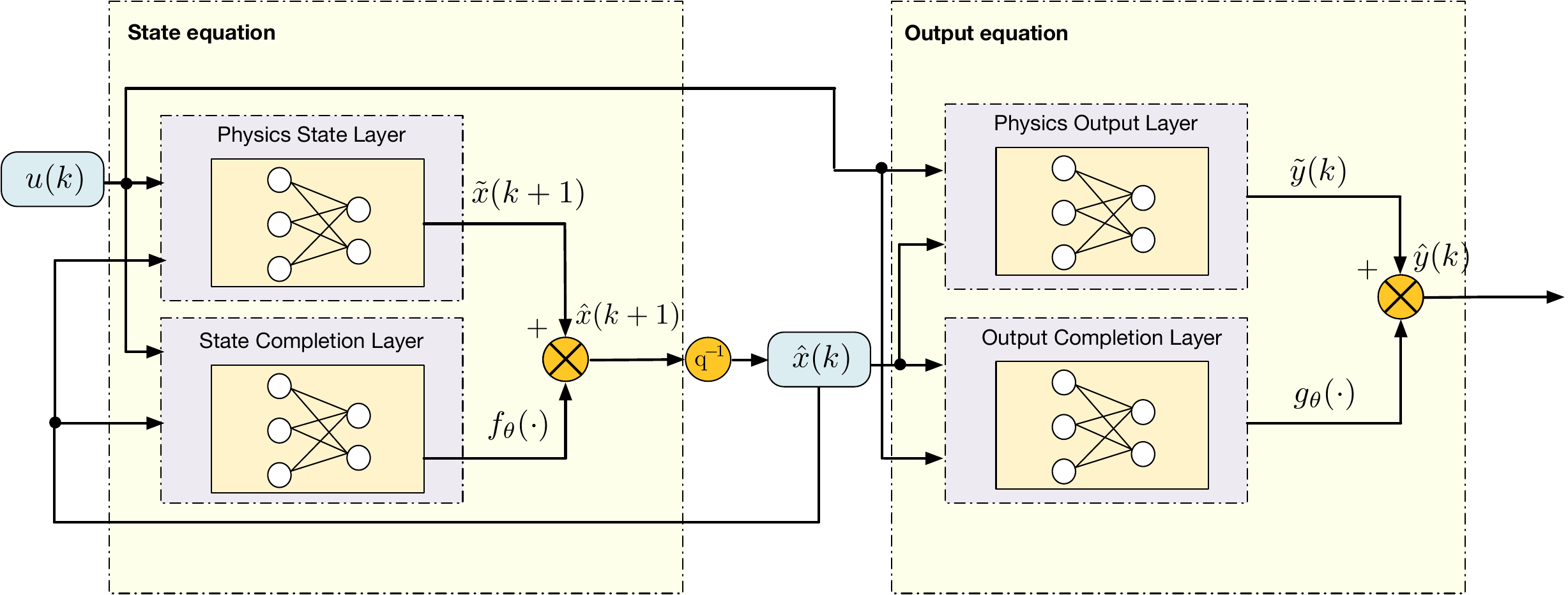}
\caption{Model structure of the physics-based SS-NN.
}
\label{fig:2}
\end{figure*}

The cost function for the proposed W-PGNN is
given by:
\be
\label{cost_WPGNN}
V(\theta,N,\bar{N}) =V_{\mathrm{Data}}(\theta,N) +  V_{\mathrm{Reg}}(\theta,w,\bar{N})
\ee
where the novel weighed regularization term is given by:
\be
V_{\mathrm{Reg}}(\theta,w,\bar{N})\triangleq  \frac{1}{\bar{N}}\sum_{j=1}^{\bar{N}} w_j \left(\gamma_x e^x_j +  \gamma_y e^y_j\right)
\ee
where $e^x_j \triangleq \left\|f_\theta(\hat{\bar{x}}(j), \bar{u}(j))\right\|^2$, $e^y_j \triangleq \left\|g_\theta(\hat{\bar{x}}(j), \bar{u}(j))\right\|^2$. The weight vector $w\in\mathbb{R}^{\bar{N}\times 1}$ is defined as:
\be
\label{6}
\begin{aligned}
w_j &\triangleq \frac{1}{\sum_{k=1}^{{N}} h_k(\bar{z}_j)+\epsilon},\\
h_k(\bar{z}_j) &= \exp\left(- \frac{\|\hat{z}_k-\bar{z}_j\|^2}{2\sigma^2}\right),
\end{aligned}
\ee
with the state-input pairs $\hat{z}_k=(\hat{x}(k),{u}(k))$ and $\bar{z}_j=(\hat{\bar{x}}(j),\bar{u}(j))$. Furthermore, $\hat{x}(k)$ and $\hat{\bar{x}}(j)$ denote the responses of the estimated model \eqref{2b} to the training input $u$ and regularization input $\bar{u}$, respectively.
$\sigma$ represents the center width of $h_k(\cdot)$, and $\epsilon$ is a small constant. Additionally, it is clear that if the weight $w_j$ is set to $1$ for all $j=1,...,\bar{N}$, then it is a classical PGNN with state-level regularization; if the weight $w_j$ is set to $0$ for all $j=1,...,\bar{N}$, then the cost function \eqref{cost_WPGNN} will completely fall back to the baseline \eqref{cost_ANN}.

One can observe from \eqref{6} that, the further away the current regularization state-input pair $(\hat{\bar{x}}(j),\bar{u}(j))$ is from the training dataset, the larger the cost $V_{\mathrm{Reg}}(\theta,w,\bar{N})$ will be. This pushes $f_\theta$ and $g_\theta$ towards zero, consequently bringing the identified model closer to the prior model in that region of the joint input-state space.
Additionally, the terms $e^x_j$ and $e^y_j$ share the same weight because both the state and the output function estimate depend on the same state-input pair.  
Note that the proposed cost function does not penalize the difference between the output of the a priori known model and the estimated model, but it rather penalizes the difference between both the state and output function of both models in regions where little information is provided by the measured data. As a consequence, the regularization state-input pair $(\hat{\bar{x}}(j),\bar{u}(j))$ should cover the full intended range of operation of the completed model for an effective regularized model augmentation through the proposed approach.

\subsection{Implementation}
\label{subsec:4-2}
The whole structure of the physics-based SS-NN is illustrated in Fig.~\ref{fig:2}.
Specifically, the SS-NN architecture mainly comprises two components, namely the physics state/output layer and the state/output completion layer, where the physics state/output layer is employed to represent the prior known model  $\tilde{f}(\cdot)$ and  $\tilde{g}(\cdot)$ given in \eqref{2a}, and the state/output completion layer is utilized for estimating the unknown dynamics $f_\theta(\cdot)$ and  $g_\theta(\cdot)$ given in \eqref{2b}. \YLMS{It is worth mentioning that the prior model \eqref{2a} should have the same state dimension as the estimated model \eqref{2b}, which is a limitation of the proposed approach.}

\textbf{Training}: \YLMS{The hyperparameters of the classical PGNN, $\gamma$, and the proposed W-PGNN, $\gamma_x$, $\gamma_y$, $\sigma$, $\epsilon$, are determined by grid searching on a validation dataset.}
\YLMS{Specifically, the selection of $\sigma$ depends on the density of data distribution, for instance, sparsely distributed data can necessitate choosing a larger $\sigma$.}
The weights and bias parameters $\theta={\mathrm{vec}}(W_x, W_{fx},W_{fu}$ $,W_y, W_{gx},W_{gu},b_f, b_x,b_g, b_y)$ of the SS-NN are trained by minimizing the cost function \eqref{cost_WPGNN} via gradient-based approaches.
Several optimization algorithms have been proposed to solve this problem, such as quasi-Newton \citep{fletcher1963rapidly} and conjugate gradients \citep{fletcher1964function} methods. In this paper, the Levenberg-Marquardt algorithm \citep{levenberg1944method} is employed to find the minimum of \eqref{cost_WPGNN}. All the algorithms are implemented in the Matlab Deep Learning Toolbox and the Matlab Optimization Toolbox.

\textbf{Initialization of model completion layer}: Due to the use of the Levenberg-Marquardt optimization algorithm, an initial guess of the parameter values is required. 
We adopt the method in \cite{schoukens2021improved} to intuitively initialize the weight and bias parameters of the model completion layer, i.e., an explicit linear approximation is introduced:
\be
\begin{aligned}
f_\theta(x(k),u&(k)) = A_\theta x(k) + B_\theta u(k) \\
&+ \tilde{W}_x \phi\left(\left[\tilde{W}_{f x}~ \tilde{W}_{f u}\right]\left[\begin{array}{l}
x(k) \\
u(k)
\end{array}\right]+\tilde{b}_f\right)+\tilde{b}_x
\end{aligned}
\ee
\be
\begin{aligned}
g_\theta(x(k),u&(k)) = C_\theta x(k) + D_\theta u(k) \\
&~+ \tilde{W}_y \phi\left(\left[\tilde{W}_{g x} ~\tilde{W}_{g u}\right]\left[\begin{array}{l}
x(k) \\
u(k)
\end{array}\right]+\tilde{b}_g\right)+\tilde{b}_y
\end{aligned}
\ee
which leaves quite some flexibility in initializing the weights and biases of the nonlinear layers. Then, the weights and biases of the linear layer are initialized as $A_\theta=B_\theta=C_\theta=D_\theta=0$ and $\tilde{b}_x=\tilde{b}_y=0$. Additionally, the weights and biases of the nonlinear layer are initialized as $\tilde{W}_x=\tilde{W}_y=0$, and  $\tilde{W}_{fx},\tilde{W}_{fu},\tilde{W}_{gx},\tilde{W}_{gu}, \tilde{b}_f, \tilde{b}_g$ are randomly initialized by $\mathcal{U}(-1,1)$. This chosen parameter initialization ensures that the initial model behaves like the a priori provided physics model. During the optimization, the weights $\tilde{W}_x=\tilde{W}_y$ will become nonzero, and this will activate the model completion part of the model.

\section{Simulation Study}
\label{sec:5}
In this section, simulation results are presented to illustrate the effectiveness of our proposed W-PGNN approach. A 1-D example is conducted to validate the superior learning performance of the proposed W-PGNN approach over the baseline and classical PGNN approaches. Consider a SISO system:
\be
\label{SISO}
\begin{aligned}
x(k+1) &= ax(k)+bu(k)+\Delta(x(k)),\\ y_0(k)&=x(k) 
\end{aligned}
\ee
with $a = 0.8187$ and $b =0.1813$. 
The function $\Delta(x(k))$ is defined as:
\be
\Delta(x)=0.2\left(e^{-\frac{x^2}{l^2}}-e^{-\frac{(x-c)^2}{l^2}}\right)
\ee
with $c=-0.3$, $l=0.2$, \YLTR{which represents the local
nonlinearity that is not able to be expressed by the given baseline physics model. Then, the augmentation structure \eqref{2b} is given in terms of the prior physics model} $\tilde{f}(x(k),u(k))=ax(k)+bu(k)$, $\tilde{g}(x(k),u(k))=x(k)$ and the completion function $f_{\theta}(x)$ aimed to identify $\Delta(x(k))$ while $g_\theta=0$  in this case.
Thus, the goal is to augment the prior model
$\tilde{f}(x(k),u(k))$ with a well-estimated $f_{\theta}$ based on the proposed W-PGNN approach.


\begin{figure}[t] 
\centering\includegraphics[width= 3in]{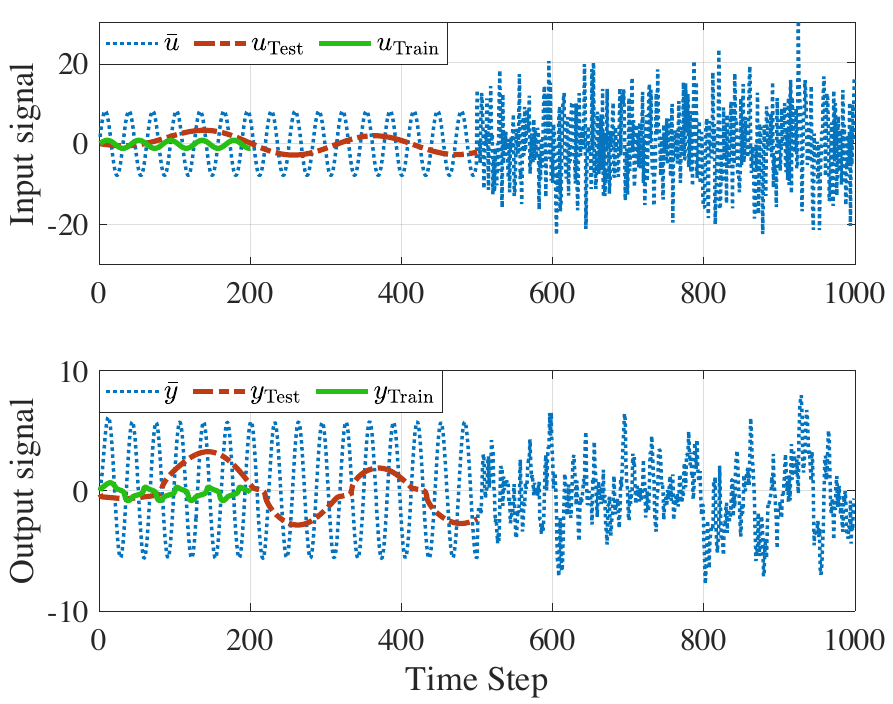}\vspace{-3mm}
\caption{Illustration of the training, regularization, and test input and output signals used in the simulation.}
\label{fig:2_5}
\end{figure}


\begin{figure}[t] 
\centering
        
            \includegraphics[width=3in]{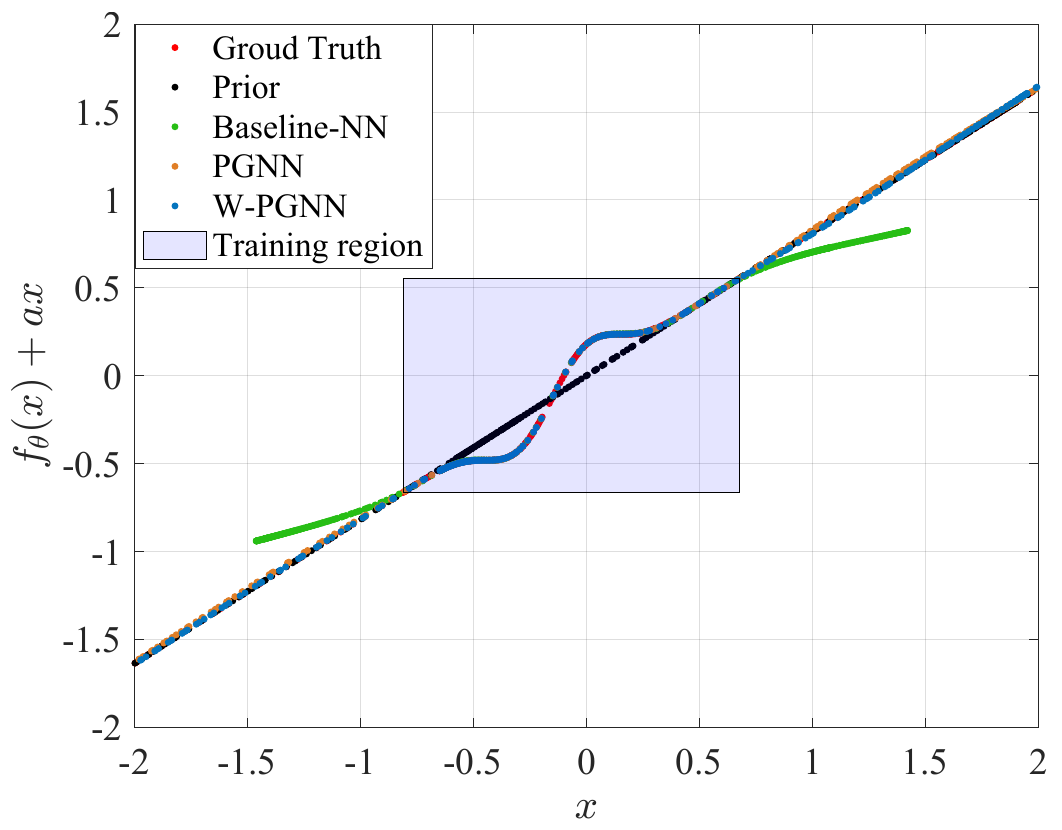}
             \vspace{-2mm}
            \includegraphics[width=3in]{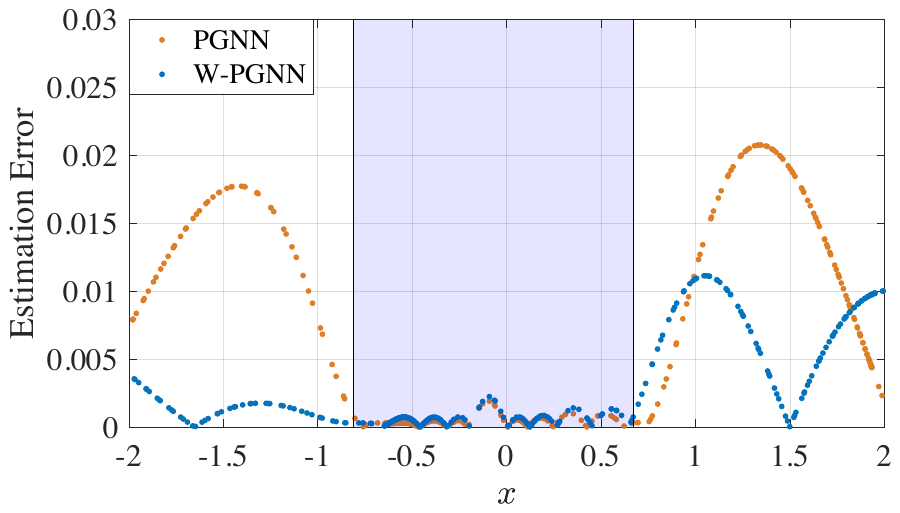}
            
        \caption{Estimation results under the considered approaches in terms of $f_\theta(x)+ax$ (top), and the absolute value of estimation error of $f_\theta(x)$ over $x$ (bottom). }\label{fig:4}
\end{figure}

With this SISO system \eqref{SISO},
the training input is selected as  $u_{\mathrm{train}}(k)=\sin(0.15k)-0.2$ with $N = 200$ samples to generate the training dataset $\mathcal{D}$.
Furthermore, the regularization input signal is designed as a concatenation of signals $\bar{u}(k)=8\sin(0.2k)$ and $\bar{u}(k)=(8+(2/500)k)\mathcal{N}(0,1)$, each with a length of 500, respectively, leading to a $\mathcal{D}_{\mathrm{Reg}}$ with size $\bar{N}=1000$.
In addition, the test input signal is selected as $u_{\mathrm{test}}(k)=\sin(0.01k+0.5)+\sin(0.02k-0.1)-2\sin(0.03k+0.2)$ with $N=500$ samples, \YL{which will explore a much larger region of input-output space than the training dataset.}  It is worth noting that only noise at the output of the system is present \YLTR{with SNR$\approx$40dB}. The aforementioned signals are visualized in Fig.~\ref{fig:2_5}, which implies that the training dataset is significantly less informative than the test dataset. This is in line with the model augmentation philosophy of this work: \YLMS{as an adequate prior model is already in place, we only would like to augment this model using a simple dataset dedicated to a particular region. }


\YLMS{To construct the NN model, the activation function is chosen as the radial basis function because of its universal approximation capability. A total of 20 neurons ($n_n=20$) are used in the state/output completion layer.} Moreover, to determine the most suitable hyperparameters for classical PGNN and the proposed W-PGNN, a grid search is conducted on the validation dataset $\mathcal{D}_{\mathrm{Val}}$, which is generated by validation input signal $u_{\mathrm{Val}}(k)=1.08\sin(0.15k)-0.2$ for the classical PGNN, and $u_{\mathrm{Val}}(k)=u_{\mathrm{Train}}(k)$ for the W-PGNN. Both of them are 500 samples long.  The results of the hyperparameter search are: $\gamma= 10^{-3}$, $\gamma_x=\gamma_y=10^{-4}$, $\sigma = \sqrt{0.001}$, and $\epsilon = 0.1$.
Then all three approaches are trained on the obtained dataset $\mathcal{D}$ and $\mathcal{D}_{\mathrm{Reg}}$, of which the parameters $\hat{\theta}$ are optimized by the Levenberg-Marquardt algorithm, as mentioned in Section \ref{subsec:4-2}.

\begin{figure}[t] 
\centering\includegraphics[width= 3in]{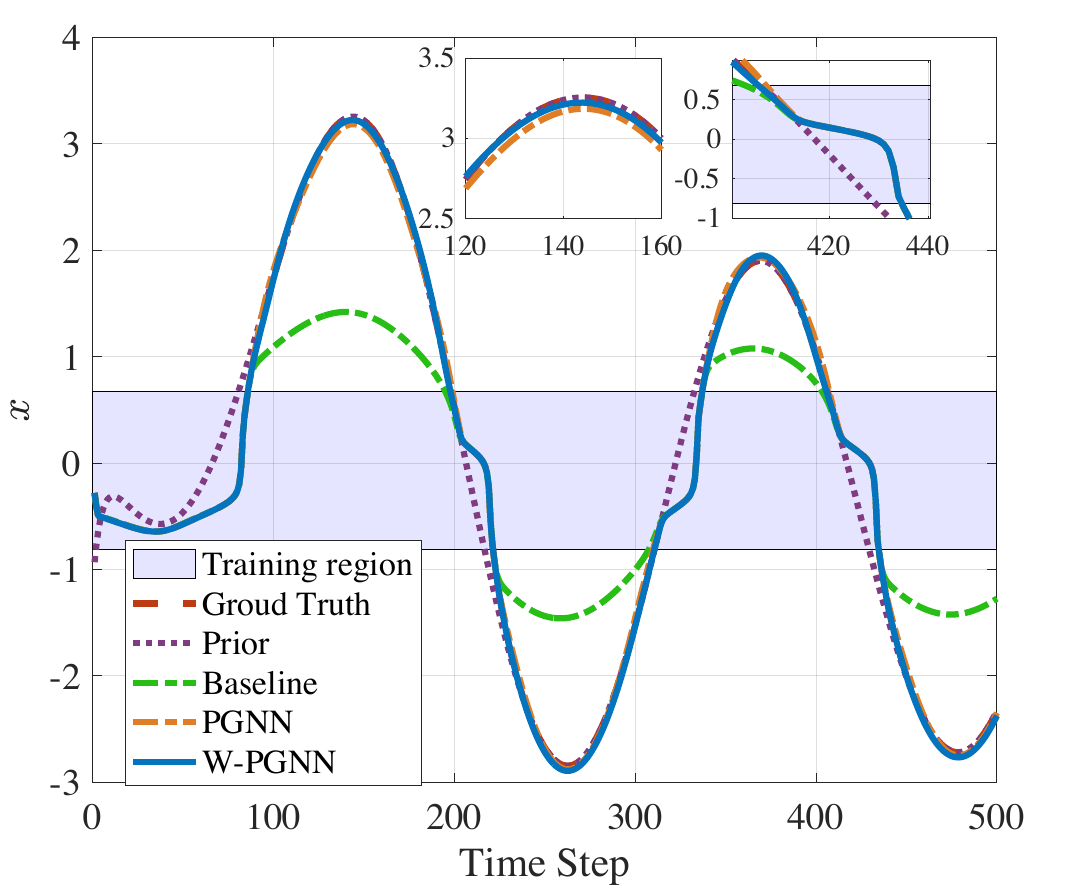}\vspace{-3mm}
\caption{Test results for the augmented state-space model by the three approaches on the test dataset. The proposed W-PGNN shows the best estimation results both within and outside the training region. }
\label{fig:5}
\end{figure}

\YLMS{The estimation results in terms of $f_\theta(x)+ax$, and the absolute value of estimation error $\Delta(x)-f_\theta(x)$ are depicted in Fig.~\ref{fig:4},} where the shaded area indicates the training data region and the black dots represent the linear prior model. It is clear that all three approaches are capable of capturing the true model well inside the training region, however, the baseline NN approach has poor generalization performance with the unseen data. Moreover, both the classical PGNN and the proposed W-PGNN approaches show good learning results outside the training region. However, the performance of the proposed W-PGNN approach is approximately 20\% better compared to the classical PGNN (see also Table~\ref{tab:1}), mainly resulting from the novel weighted-regularization physics-based term in the cost function, which enables the learned model to follow the ground truth within the range of the training data, and in turn, be forced toward the linear prior model within the low-informative data area. This can also be seen in Fig.~\ref{fig:5}, where the zoom-in sub-figures show the estimation trajectories inside and outside the training region, respectively. It can be observed that despite the test dataset being much larger than the training dataset the proposed W-PGNN still has the capability of identifying the system in the whole state space with the highest estimation accuracy.

Furthermore, a Monte Carlo simulation with 10 runs under random initial parameters is conducted to compare the estimation error of the three approaches.
To assess the simulation performance of the identified models, the following  \emph{root mean squared error} (RMSE) on the test dataset is utilized: 
\be
e_{\mathrm{RMSE}}=\sqrt{\frac{1}{N} \sum_{k=1}^N(y(k)-\hat{y}(k|\theta))^2}
\ee

\begin{table}[t]
\centering
\caption{Quantitative evaluation of the performance of the three approaches in terms of their RMSE on the training and test dataset.}
\label{tab:1}
\begin{tabular}{llll}
\toprule[1.5pt]
\multicolumn{2}{l}{\multirow{2}{*}{Approach}} & \multicolumn{2}{c}{$e_{\mathrm{RMSE}}$($\times 10^{-2}$)} \\ \cline{3-4} 
\multicolumn{2}{l}{}                          & Training set  & Test set \\ \hline
\multicolumn{2}{l}{Baseline}                  & 0.241$\pm$0.027                 & 199.7$\pm$  170.3     \\
\multicolumn{2}{l}{Classical PGNN}             & 0.206$\pm$0.024                 & 5.791$\pm$2.810       \\
\multicolumn{2}{l}{\bf{W-PGNN}}                    & 0.209$\pm$0.021                 & $\mathbf{4.303\pm1.503}$       \\ \bottomrule[1.5pt]
\end{tabular}
\end{table}

Table~\ref{tab:1} quantifies the RMSE and its variability of the three considered approaches on the training and test dataset over 10 runs. One can see that the achieved RMSE of the proposed W-PGNN significantly improves and shows better generalization performance on the unseen dataset compared to the baseline NN and classical PGNN.

\vspace{-2mm}
\section{Conclusion}
\label{sec:6}

A novel PGNN-based model completion strategy is proposed in this paper for nonlinear state-space model identification. Specifically, we enhance the interpretability and generalization performance of classical PGNN by introducing a weighted function regularization strategy, i.e., the W-PGNN. A new weighted regularization cost function is presented \YLTR{to penalize the difference between the physics and identified model at both the state and output levels in regions with low information content.}
The proposed strategy provides new perspectives into the fusion of physics-guided and black-box data-driven modeling approaches, especially in cases where the available data is limited.
The effectiveness of W-PGNN has
been analyzed and demonstrated by numerical simulations and compared with some classical ANN modeling methods. \YL{Future work will focus on extending the application scenarios of the proposed W-PGNN method to more complex and larger benchmarks.}

\bibliography{ifacconf}   

\end{document}